\documentclass[12pt]{article}
\begin{document}
\pagenumbering{arabic}
\begin{titlepage}
\title{Gravitational energy, gravitational pressure, and the thermodynamics of a charged black hole in teleparallel gravity}

\author{K. H. C. Castello-Branco$\,^{1\ast}$ \\ and
J. F. da Rocha-Neto$\,^{2\dagger}$}
\date{}
\maketitle

\begin{center}
1 Universidade Federal do Oeste do Par\'a, \\
Av. Marechal Rondon, 68040-070. Santar\'em, PA, Brazil.
\end{center}

\begin{center}
2 Instituto de F\'isica, Universidade de Bras\'ilia, \\
70910-900, Bras\'ilia, DF, Brazil.
\end{center}

\begin{abstract}

We investigate, in the case of a Reissner-Nordstr\"om black hole, the definitions of gravitational energy and gravitational 
pressure that na-\linebreak turally arise in the framework of the {\it Teleparallel Equivalent of Ge-\linebreak neral Relativity}. In particular, 
we calculate the gravitational energy enclosed by the event horizon of the black hole, $E$, and the radial pre-\linebreak ssure over it, $p$. 
With these quantities we then analyse the thermodynamic relation $dE\,+\,pdV$ (as $p$ turns out to be a density, $dV$ is 
actually given by $dV=dr d\theta d\phi$, in spherically-type coordinates). We compare the latter with the standard first law 
of black hole dynamics. Also, by identifying $TdS = dE\,+\,pdV$, we comment on a possible modification of the standard, 
Bekenstein-Hawking entropy-area relation due to gravitational energy and gravitational pressure of the black hole. The 
infinitesimal variations in question refer to the Penrose process for a Reissner-Nordstr\"om black hole.

\end{abstract}



{\footnotesize
\noindent $\ast$ khccb@yahoo.com.br\\
\noindent $\dagger$ rocha@fis.unb.br}

\end{titlepage}

\bigskip

\section{Introduction}

$ $ 

As is well-known the behaviour of black holes as thermodynamic systems deeply connects gravitation, 
quantum mechanics and thermodynamics. The black hole surface gravity $\kappa$ 
plays the role of temperature, its horizon area $A$ that of entropy, and its mass 
$M$ that of internal energy. This striking connection initially flourished from a 
close analogy between the laws of black hole dynamics and the laws of thermodynamics. 
It was only later that it was put on a firm basis, due to the discovery of Hawking 
that quantum mechanical effects permit a black hole to create and emit particles like a 
hot body with temperature $\kappa /2\pi$ (in units with $G=c=\hbar =\kappa_{B}=1$) \cite{Hawking}. 
Nevertheless, such connection is considered to be still poorly understood presently \cite{Poisson}. 

\bigskip

It is also known that a notion of gravitational energy can be ascribed to black holes, not to mention 
the gravitational energy transported by gravitational waves. By means, for instance, of the quasilocal
energy approach of Brown and York, which is based on a Hamilton-Jacobi formulation of general relativity, 
one can compute the gravitational energy enclosed by the event horizon of a black hole \cite{Brown-York}. 
We note that the old attempts to define gravitational energy by means of pseudo-tensors are not appropriate, 
as well as definitions based on space-time symmetries (see, for instance, item (1) in Introduction of Ref. 
\cite{Mann-2012}). Since a black hole encloses gravitational energy, one can then naturally consider that 
such energy plays a role on the thermodynamical behaviour of black holes, as internal energy (see Ref. 
\cite{Brown-York}). 
     
\bigskip

The notion of gravitational energy has also been shown to be well-defined in the framework of the 
\textit{Teleparallel Equivalent of General Relativity} (TEGR). The TEGR  \cite{Hehl1,Nester1989,Maluf1994,Hehl2,Maluf4,Obukov-Pereira,Obukov-Rub,Nester,Maluf-CQG2007,
Maluf-Ulhoa2008,CastelloBr-Rocha2012,Aldrovandi-JG} is not a new theory of gra-\linebreak vity, but an 
alternative geometric formulation of general relativity, which (in its simplest formulation) has as basic field 
variables only \textit{tetrad fields}. The space-time of the theory is endowed only with torsion, rather than curvature. 
In this setting it is then possible to define a distant parallelism or \textit{telepara-\linebreak llelism} of vectors at different points of 
space-time, provided that they have identical components with respect to the local tetrads 
at the points consi-\linebreak dered. The equivalence of the theory with general relativity is at the 
level of field equations \cite{Maluf1994}. For a recent review on TEGR we refer the reader to Ref. \cite{ReviewTEGR-Maluf}. 
In the TEGR the notion of gravitational energy, $E$, has been defined from 
the Hamiltonian formulation of the theory \cite{Maluf4} and latter it has been shown that it derives directly from 
the field equations of the theory \cite{Maluf1}. Recently, the notion of \textit{gravitational pressure}, $p$, 
over the event horizon of a black hole has also been shown to be well-defined in the realm of the 
TEGR. The gravitational pressure naturally arises from the field equations of the TEGR and from the gravitational 
energy-momentum tensor defined in the theory. The spatial components of such energy-momentum tensor yield 
the standard definition of the gravitational pressure in the TEGR \cite{maluf-pressao-Kerr} (see also \cite{Maluf1}, 
in which the definition was first established). On Sec. 2, we review such definitions of gravitational energy and 
gravitational pressure. 

\bigskip

In this work, by considering a Reissner-Nordstr\"om black hole, we further extend the investigation of the concept 
of gravitational pressure which arises in the context of the TEGR and that has recently been studied 
in the case of a Kerr black hole \cite{maluf-pressao-Kerr}. It is important to better understand the nature of the 
gravitational pressure and its effects on the thermodynamic behaviour of black holes. One of our main goals is to 
make the comparison of the relation $dE\,+\,pdV$, obtained entirely in the TEGR, with the standard first law of 
black hole mechanics. The variations in the latter quantity are considered to be related to the 
Penrose process \footnote{As one knows, the Penrose process occurs not only with rotating black holes, 
but also with a charged static black hole. In the case with rotation, if a particle with non-zero angular momentum 
has negative energy inside the ergosphere of a Kerr black hole, then an extraction of energy and angular momentum 
from the black hole will take place. Although no ergoregion like that of the Kerr case exists for a Reissner-Nordstr\"om 
black hole, there is something like it, since it is possible for a particle to arrive at the horizon with 
negative energy, provided its electric charge is opposite to that of the black hole. If such a particle falls down 
into the black hole, this process will lead to an extraction of mass and electric charge from the black hole \cite{Chris-Ruff}. 
The extracted energy comes at the expense of some of the mass and charge of the black hole.}. 
The variation $dV$ is basically obtained by means of the variation of the radius of the event 
horizon, $r_{+}$, when the parameters $M$ (mass) and $Q$ (charge) of the black hole vary by infinitesimal 
amounts $dM$ and $dQ$ (as $p$ turns out to be a density, then $dV$ is actually given by $dV=dr d\theta d\phi$). 
Analogously to Ref. \cite{maluf-pressao-Kerr}, in which a Kerr black hole was considered, we remark that our analysis 
is essentially restricted to the event horizon of a Reissner-Nordstr\"om black hole, without considering any property 
of its horizon area $A$. It is only after we derive, entirely in the framework of the TEGR, our main result, 
which is the quantity $dE\,+\,pdV$, is that, in order to compare it with the standard first law of black hole mechanics, namely

\begin{equation}
\frac{\kappa}{8\pi}dA = dM - \Phi_{H}\,dQ\,,
\label{1a-lei-mec-bn}
\end{equation}
in which $\Phi_{H}=Q/r_{+}$ is the electrostatic potential at $r_{+}$, we will consider the area $A$ and 
its property according to which by no continuous process can it be decreased (\textit{i.e.}, 
$dA \geq 0$), the latter being, as is well-know, simply a consequence of the fact 
that the irreducible mass of a black hole cannot be decreased by any continuous process, as a Penrose process 
\cite{Chris,Chris-Ruff}. We note that, as far as we know, the concept of gravitational pressure in the first 
law of black hole (thermo-)dynamics (for static, spherically symmetric black holes) has been firstly introduced 
by Brown and York \cite{Brown-York}, who defined a \textit{surface} pressure, whereas recently the use of the 
concept of gravitational pressure has been made by Dolan \cite{Dolan}, by considering that the cosmological 
constant plays the role of pressure.

\bigskip

Another important question is how the gravitational pressure affects the efficiency of the Penrose process. 
A comparison of the effect of the gravitational pressure on the 
efficiency of the Penrose process for a Kerr black hole (obtained in Ref. \cite{maluf-pressao-Kerr}) 
with that for a Reissner-Nordstr\"om black hole (which we investigate in this paper) may be important 
in order to achieve a better understanding of the concept of gravitational pressure. For the Kerr case, 
according to Ref. \cite{maluf-pressao-Kerr}, it is shown that the efficiency of the Penrose process 
in the context of the TEGR is lower than in the ordinary thermodynamic formulation in general relativity. 

\bigskip

\section{Gravitational energy-momentum and gra-\\
vitational pressure in the TEGR}

$ $ 

The equivalence of the TEGR with Einstein's general relativity is obtained by means of an identity
 that relates the scalar curvature $R(e)$ constructed out of the tetrad field and a combination 
of quadratic terms of the torsion tensor \cite{Hehl1, Hehl2, Blago, Ortin}
 
\begin{equation}
eR(e) \equiv -e\left({1\over 4}T^{abc}T_{abc} + {1\over 2}T^{abc}T_{bac}  - T^{a}T_{a}\right)
+ 2\partial_{\mu}(eT^{\mu}),
\label{equi}
\end{equation}
where $e = det(e^{a}\,_{\mu})$,  $T_a=T^b\,_{ba}\,$, $T_{abc}=e_b\,^\mu e_c\,^\nu T_{a\mu\nu}\,$ and
$T_{a\mu\nu}$  is the torsion tensor, defined by
$T_{a\mu\nu} = {\partial}_{\mu}e_{a\nu}-{\partial}_{\nu}e_{a\mu}\,$.

\bigskip

In the framework of the TEGR the lagrangian density is given in terms of the combinations of the
quadratic terms in the equation above, {\it i.e.},
\begin{eqnarray}
L&=& -k e\left(\frac{1}{4}T^{abc}T_{abc}+\frac{1}{2}T^{abc}T_{bac}-
T^aT_a\right) - \frac{1}{c}L_m\nonumber \\
&\equiv& -ke\Sigma^{abc}T_{abc}-\frac{1}{c}L_m\,, 
\label{lagrang-tegr}
\end{eqnarray}
in which $k=c^{3}/16\pi G\,$, and $\Sigma^{abc}$ is defined by

\begin{equation}
\Sigma^{abc}= \frac{1}{4} \left(T^{abc}+T^{bac}-T^{cab}\right)
+\frac{1}{2}(\eta^{ac}T^b-\eta^{ab}T^c)\,,
\label{def-sigma}
\end{equation}
and $L_m$  is the Lagrangian density for matter fields. 
 
\bigskip
 
The field equations derived from (\ref{lagrang-tegr}) for the tetrad field is equivalent to Einstein's 
equations, and it reads

\begin{equation}
e_{a\lambda}e_{b\mu}\partial_\nu (e\Sigma^{b\lambda \nu} )-
e \left(\Sigma^{b\nu}\,_aT_{b\nu\mu}-
\frac{1}{4}e_{a\mu}T_{bcd}\Sigma^{bcd} \right)=\frac{1}{4kc}eT_{a\mu}\,,
\label{eq-campo-tegr}
\end{equation}
in which
$eT_{a\mu}=\delta L_m / \delta e^{a\mu}$. 
In fact, one can show that the left-hand side of the latter equation 
may be written exactly as $\frac{1}{2}e\left[ R_{a\mu}(e)-\frac{1}{2}e_{a\mu}R(e)\right]$. Therefore 
it turns out that (\ref{eq-campo-tegr}) is the Einstein's equations of general relativity in terms of 
tetrad fields. From now on we will set $G=c=1$, unless we say otherwise.

\bigskip

As shown in Ref. \cite{Maluf1}, Eq. (\ref{eq-campo-tegr}) may be simplified as 

\begin{equation}
\partial_\nu(e\Sigma^{a\lambda\nu})=\frac{1}{4k}
e\, e^a\,_\mu( t^{\lambda \mu} + T^{\lambda \mu})\;,
\label{5}
\end{equation}
where $T^{\lambda\mu}=e_a\,^{\lambda}T^{a\mu}$ and
$t^{\lambda\mu}$ is defined by

\begin{equation}
t^{\lambda \mu}=k\left(4\Sigma^{bc\lambda}T_{bc}\,^\mu-
g^{\lambda \mu}\Sigma^{bcd}T_{bcd}\right)\,.
\label{6}
\end{equation}
In view of the property 
$\Sigma^{a\mu\nu}=-\Sigma^{a\nu\mu}$ it follows that

\begin{equation}
\partial_\lambda
\left[e\, e^a\,_\mu( t^{\lambda \mu} + T^{\lambda \mu})\right]=0\,.
\label{7}
\end{equation}
This equation then yields the following \textit{continuity (or balance) equation},

\begin{equation}
\frac{d}{dt} \int_V d^3x\,e\,e^a\,_\mu (t^{0\mu} +T^{0\mu})
=-\oint_S dS_j\,
\left[e\,e^a\,_\mu (t^{j\mu} +T^{j\mu})\right]\,.
\label{8}
\end{equation}
Thus $t^{\lambda\mu}$ can be identified as the {\it gravitational energy-momentum tensor} 
\cite{Maluf1, Maluf2} \footnote{We note that a pseudo-tensor for gravitational energy-momentum 
in the realm of the TEGR was proposed in Ref. \cite{JG-PRL2000}, but which is different from our 
Eq. (\ref{6}). The mentioned expression of Ref. \cite{JG-PRL2000} is shown therein to be equivalent to 
the M\"oller's pseudo-tensor expression in his formulation of gravity by means of tetrad fields \cite{Moller}.}, 

\begin{equation}
P^a=\int_V d^3x\,e\,e^a\,_\mu (t^{0\mu} 
+T^{0\mu}),
\label{9}
\end{equation}
as the total energy-momentum contained within a volume $V$ of the three-dimensional space,

\begin{equation}
\Phi^{a}_{g} = \oint dS_{j}(e e^{a}\,_{\mu}t^{j\mu}),
\label{fluxog}
\end{equation}
as the energy-momentum flux of the gravitational field and

\begin{equation}
\Phi^{a}_{m} = \oint dS_{j}(e e^{a}\,_{\mu}T^{j\mu}), 
\label{fluxom}
\end{equation}
as the energy-momentum flux of matter.

\bigskip

In view of Eq. (\ref{5}), the Eq. (\ref{9}) may be written simply as 

\begin{equation}
P^a=-\int_V d^3x \partial_i \Pi^{ai}\quad,
\label{def-energia-mom}
\end{equation}
where $\Pi^{ai}=-4ke\,\Sigma^{a0i}$ is the momentum
canonically conjugated to $e_{ai}$. This expression was first obtained in the 
context of Hamiltonian formulation of the TEGR in vacuum (see Ref. \cite{Maluf5}).  It is invariant under
coordinate transformations of the three-dimensional space and under time reparametrizations. The gravitational energy 
enclosed by a three-dimensional volume, limited by a surface $S$, is defined by the $a=(0)$ 
component of Eq. (\ref{def-energia-mom}), {\it i.e.},
\begin{equation}
P^{(0)}=  \oint_{S}\, dS_{i}\,4k e\Sigma^{(0)0i}\,.
\label{energia-grav}
\end{equation}
This definition has been succesfully applied to several important space-times, as for determining 
the energy enclosed by the event horizon of a Kerr black hole \cite{Maluf4}, the energy (mass) loss 
described by the Bondi metric \cite{Maluf-Faria2004} and the energy of gravitational waves 
\cite{Maluf-Ulhoa2008}, for instance.

\bigskip

Let us now see how pressure naturally arrises from some of the latter equations. 
It follows from Eq.'s (\ref{5}), (\ref{8}) and (\ref{9}) that

\begin{equation}
{{dP^a}\over {dt}}=
-4k\oint_S dS_j\,
\partial_\nu(e\Sigma^{a j\nu})\,.
\label{17}
\end{equation}
If one now makes the Lorentz index $a$ to be restricted to $a=(i)=(1),(2),(3)$, then Eq. (\ref{17}) can be 
written as

\begin{equation}
{{dP^{(i)}}\over {dt}}= \oint_S dS_j\, (-\phi^{(i)j})\,,
\label{18}
\end{equation}
in which

\begin{equation}
\phi^{(i)j}=4k\partial_\nu(e\Sigma^{(i)j\nu}) \,.
\label{19}
\end{equation}
We note that Eq. (\ref{18}) is precisely the Eq. (39) presented  in \cite{Maluf1}.
As remarked by Maluf in Ref. \cite{Maluf1}, the left-hand side of Eq. (\ref{18}) represents the momentum 
divided by time, what implies it has the dimension of force. And since on the right-hand side of 
Eq. (\ref{18}) $dS_j$ is an element of area, one sees that $-\phi^{(i)j}$ can be understood as force per unit 
area, \textit{i.e.}, a \textit{pressure density}; it represents the pressure along the $(i)$-direction over an element 
of area oriented along the $j$-direction. If one considers, for instance, cartesian coordinates, then the 
index $j=1,2,3$ represents the directions $x,y,z\,$, respectively. To compute the radial pressure over the event horizon of a black hole, in spherical-type coordinates, we set 
$j = r,  \theta, \varphi$. In this case we need to consider only the index $j = 1$, which 
is associated with the radial direction. Therefore, in spherical-type 
coordinates the density $\phi^{(r)1}$ is given by
\begin{equation}
-\phi^{(r)1} = -(\sin\theta\cos\varphi \phi^{(1)1}+\sin\theta\sin\varphi\phi^{(2)1}+\cos\theta\phi^{(3)1}),
\label{19.1}
\end{equation}
from which we define the \textit{radial pressure} $p$ as
\begin{equation}
p(r) = \int^{2\pi}_{0}d\varphi\int^{\pi}_{0}d\theta[-\phi^{(r)1}]\,.
\label{19.2}
\end{equation}

\bigskip

In the next two sections we will compute both the gravitational energy enclosed by the event horizon 
of a Reissner-Nordstr\"om black hole and the radial pressure over its surface.

\bigskip 

\section{Gravitational energy of a Reissner-Nordstr\"om black hole}

$ $ 

In standard spherical-type coordinates the line-element for a Reissner-Nordstr\"om black hole is given by

\begin{equation}
ds^{2} = -\alpha ^{2}dt^{2}+ \alpha ^{-2}dr^{2}+r^{2}\left (d\theta^{2}+ \sin ^{2}\theta d\varphi^{2}\right )\,,
\label{metrica} 
\end{equation}
in which 

\begin{equation}
\alpha =\left (1-\frac{2M}{r}+\frac{Q^{2}}{r^{2}}\right )^{1/2}\,.
\label{funcao-metrica} 
\end{equation}
The parameters $M$ and $Q$ are the mass and charge of the black hole, in geometrized units, respectively. 
The roots of $\alpha = 0$ are 

\begin{equation}
r_{\pm}= M \pm \sqrt{M^{2}-Q^{2}}\,,
\label{horizons} 
\end{equation}
with $r_+$ and $r_-$ being the radius of the (external) event horizon and the (internal) Cauchy horizon, 
respectively.

\bigskip

Let us now choose a set of tetrad fields related to (\ref{metrica}). Tetrad fields, which are the basic field 
variables of the TEGR, can naturally be interpreted as {\it reference frames} 
adapted to observers in spacetime \cite{Hehl}, an interpretation that has been explored in investigations on both the 
energy and angular momentum of the gravitational field in TEGR \cite{Maluf-CQG2007}. To each observer in spacetime one can adapt 
a tetrad field in the following way \cite{Hehl}. If $x^{\mu}(s)$ denotes the world line $C$ of an observer in spacetime, 
where $s$ is the observer's proper time, the 
observer's four-velocity along $C$, defined by $u^{\mu}(s)=dx^{\mu}/ds$, is identified with the $a=(0)$ component 
of $e_{a}\,^{\mu}$, that is, $u^{\mu}(s)=e_{(0)}\,^{\mu}$ along $C$. In this way, each set of tetrad fields defines 
a class of referance frames in spacetime \cite{Hehl}. In what follows we will consider a set of tetrad fields adapted 
to a static observer in spacetime \cite{Maluf-CQG2007}. Given a metric $g_{\mu\nu}$, 
the tetrad field related to it can be easily obtained through $g_{\mu\nu}=\eta^{ab}e_{b\mu}e_{a\nu}$. 
The realization of tetrad fields adaptad to static observers is achieved by imposing on $e_{a\mu}$ 
the following conditions: (i) $e_{(0)}\,^{i}=0\,$, which implies that $e_{(k)0}=0\,$, and (ii) $e_{(0){i}}=0\,$, 
which implies that $e_{(k)}\,^{0}=0\,$. While the physical meaning of condition (i) is straightforward 
(the translational velocity of the observer is null, \textit{i.e.}, the three components of the frame velocity in the 
three-dimensional space are null), for condition (ii) it is not so. The latter is a condition on the 
rotational state of motion of the observer. It implies that the observer (more precisely the
three spatial axes of the observer's local spatial frame) is (are) not rotating with respect to a 
nonrotating frame (for details, we refeer the reader to Ref. \cite{Maluf-CQG2007} and references therein). Therefore, 
conditions (i) and (ii) are six conditions one can impose on the tetrad field in order to completely fix its structure. 

\bigskip

By applying above-mentioned conditions (i) and (ii), one can easily construct the set of tetrad fields related 
to (\ref{metrica}) and which corresponds to static observers (we note that for this class 
of observers the components of $T_{\mu\nu}$ that correspond to the magnetic field vanish). It is given by

\begin{equation}
e_{a\mu} = \left(\begin{array}{cccc}
-\alpha & 0 & 0 & 0\\
0  & \alpha^{-1}\sin\theta\cos\varphi & r\cos\theta\cos\varphi & -r\sin\theta\sin\varphi \\
0  & \alpha^{-1}\sin\theta\sin\varphi & r\cos\theta\sin\varphi & r\sin\theta\cos\varphi \\
0  & \alpha^{-1}\cos\theta         & - r\sin\theta       & 0
\label{tetrad}
\end{array}
\right).
\end{equation}

\bigskip

From Eq. (\ref{energia-grav}), the energy enclosed by a spherical surface of fixed radius $r$ is given by
\begin{equation}
P^{(0)}=  4k\int d\theta d\varphi\,e\Sigma^{(0)01}\,.
\label{energia-grav1}
\end{equation}

\bigskip

In order to evaluate the quantity $\Sigma^{(0)01}$, we resort to Eq. (\ref{def-sigma}). 
After a somewhat long but straightforward algebra it yields 

\begin{equation}
\Sigma^{(0)01}=\frac{1}{2}\alpha\,g^{00}g^{11}(g^{22}T_{212}+g^{33}T_{313})\,.
\label{4.3}
\end{equation}

\bigskip

The computation of the components of the torsion tensor in the latter expression is straightforward. They read

\begin{eqnarray}
T_{212}&=&-r\left(1 - \frac{2M}{r}+\frac{Q^{2}}{r^{2}} \right)^{-1/2} + \,r\,,\nonumber\\
T_{313}&=&-r\sin^2\theta  \left[\left(1 - \frac{2M}{r}+\frac{Q^{2}}{r^{2}}\right)^{-1/2} - 1\right]\,.
\label{4.4}
\end{eqnarray}

\bigskip

Now, inserting the determinant $e = r^{2}\sin\theta$ and Eqs. (\ref{4.4}) into Eq. (\ref{4.3}) we obtain

\begin{equation}
e\Sigma^{(0)01}=r\sin\theta\,\left[\frac{1}{2}
-\left(1-\frac{2M}{r}+\frac{Q^{2}}{r^{2}}\right)^{1/2}\right]+\,\frac{1}{2}r\sin\theta\,.
\label{4.5}
\end{equation} 

\bigskip

From Eq. (\ref{energia-grav1}) the energy enclosed by a spherical surface of constant radius $r$ is then given by
\begin{equation}
E (r)\equiv P^{(0)} = r \left [1-\sqrt{1-\frac{2M}{r}+\frac{Q^{2}}{r^{2}}}\,\right]\,. 
\label{energia-RN}
\end{equation}
This is precisely the expression that is obtained by means of the quasi-local energy approach 
of Brown and York \cite{york-RN}. For $Q=0$, Eq. (\ref{energia-RN}) gives the distribution of 
gravitational energy in the space-time of a Schwarzschild black hole.

\bigskip

From Eq. (\ref{energia-RN}) it follows that the energy enclosed by the event horizon of a 
Reissner-Nordstr\"om black hole is simply given by

\begin{equation}
E\equiv E(r_{+})\, =\, r_{+}\,\,. 
\label{energia-hor-RN}
\end{equation}
It is interesting to express the result (\ref{energia-hor-RN}) in terms of the \textit{irreducible mass}, $M_{irr}$, of 
the black hole \cite{Chris,Chris-Ruff}. When a charged or rotating black hole is subject 
to the Penrose process, this leads to changes in its mass and charge or its mass and angular momentum, respectively [for 
a general stationary (\textit{i.e.}, Kerr-Newman) black hole, the Penrose process will lead to the extraction of charge as well 
as angular momentum from the black hole]. In any case, the Penrose process is such that it cannot make the initial mass $M$ less than $M_{irr}$. 
For a Reissner-Nordstr\"om  black hole, the irreducible mass is given by $M_{irr}=(1/2)\,r_{+}$. Hence, from 
Eq. (\ref{energia-hor-RN}), one sees that the gravitational energy inside the event horizon of a Reissner-Nordstr\"om 
black hole can be simply written as

\begin{equation}
E = 2M_{irr}\,\,. 
\label{energia-hor-RN-Mirr}
\end{equation} 
For a Schwarzschild black hole, one simply has $M_{irr}=M$, what corresponds to the fact that there is 
neither electric nor rotational energy to be extracted from the black hole in this case. We remark the 
fact that for a Kerr black hole the gravitational energy inside its event horizon is strikingly close 
to the value $2M_{irr}$, as computed in the framework of the TEGR \cite{Maluf4}. We stress that 
this has been shown to be valid to any value of the rotation parameter. On the other hand, by applying 
the Brown-York quasilocal approach, Martinez \cite{Martinez} has shown that the gravitational energy enclosed by the horizon is 
given by $2M_{irr}$, in the regime of \textit{slow-rotation}. He conjectured that this would hold for any 
value of the rotation parameter. However, by means of a generalization of the quasilocal 
method of Brown and York, Deghani and Mann have numerically shown that such a conjecture is not valid \cite{Deghani-Mann}. 
As far as we know, the computation of the energy enclosed by the event horizon, for any value of the rotation 
parameter, via the original quasilocal approach of Brown and York has not been performed. This is due to the 
thecnical difficulty in applying it for any regime of rotation \cite{Martinez}. 

\bigskip

The results obtained in the context of the TEGR suggest that one consi-\linebreak ders the case of the Kerr-Newman 
black hole in order to see if the value $2M_{irr}$ still holds for the energy enclosed by the event horizon 
of such a black hole. In what concerns the use of the Brown-York method, it has been shown that, in the \textit{slow-rotation} 
approximation, such a value still holds for a Kerr-Newman black hole \cite{Bose}. The same result has been found by a 
computation done in the framework of the TEGR \cite{Xu-Jing}. Anyway, for the case of a Kerr-Newman black hole 
this issue deserves to be further investigated in the TEGR itself \cite{energ-KN-future}. 
  
\bigskip

\section{Radial pressure over the event horizon of a Reissner-Nordstr\"om black hole}

$ $ 

In order to evaluate the radial pressure over the event horizon of a Reissner-Nordistr\"om black hole, 
we need to compute the  conponents of $\phi^{(i)1}$ (see Eq. (\ref{19.1})). After a long but straightfoward calculation we 
obtain, considering the tetrad field given by Eq. (\ref{tetrad}), that

\begin{eqnarray}
\phi^{(1)1} &=& 4k\sin^{2}\theta\cos\varphi(\alpha\alpha'r + \alpha^2 - \alpha),\nonumber\\
\phi^{(2)1} & = & 4k\sin^{2}\theta\sin\varphi(\alpha\alpha'r + \alpha^2 - \alpha),\nonumber\\
\phi^{(3)1} & = & 4k\sin\theta\cos\theta\sin\varphi(\alpha\alpha'r + \alpha^2 - \alpha).
\label{fluxo123}
\end{eqnarray}
By inserting now the above relations into Eq. (\ref{19.1}) and performing the integration in Eq. (\ref{19.2}) 
we obtain that the radial pressure over a space-like spherical surface of radius $r$ in the space-time of a 
Reissner-Nordstr\"om black hole is given by

\begin{equation}
p(r) = -(r\alpha\alpha' + \alpha^2 - \alpha)\,,
\label{pres-rad}
\end{equation}
in which the prime denotes the derivative with respect to $r$. By making use of Eq. (\ref{funcao-metrica}) 
into Eq. (\ref{pres-rad}) one obtains 

\begin{equation}
p(r) = \frac{M}{r}+\left(1-\frac{2M}{r}+\frac{Q^{2}}{r^{2}}\right)^{1/2}-1\,,
\label{pres-rad-RN}
\end{equation}
from which follows that the radial pressure over the event horizon ($r=r_{+}$) of a 
Reissner-Nordstr\"om black hole is given by

\begin{equation}
p\equiv p(r_{+}) = \frac{M}{r_{+}}-1\,= -{(M^{2} - Q^{2})^{1/2}\over r_{+}}\,.
\label{pres-rad-hor-RN}
\end{equation}
In particular, for $Q=0$, $r_{+}$ reduces to $2M$ and one thus is left with $p = - 1/2$ (or $p = - c^{3}/2G$, by 
restoring the physical constants), which is precisely the value of the radial pressure over the horizon 
of a Schwarzschild black hole, a result that has recently been obtained by Maluf \textit{et al.} in 
Ref. \cite{maluf-pressao-Kerr}. 

\bigskip

It is instructive to compare the magnitude of the pressure over the event horizons of 
a Schwarzschild and a Reissner-Nordstr\"om black holes. As the radius of the event horizon of a 
Reissner-Nordstr\"om black hole is less than for a Schwarzschild one, \textit{i.e.}, 
$(r_{+})_{RN}\,<\,(r_{+})_{Sch}$, from Eq. (\ref{pres-rad-hor-RN}) one sees that the pressure 
over the event horizon of a Reissner-Nordstr\"om black hole is, \textit{in modulus}, greater than 
for a Schwarzschild one. This is physically reasonable, since a Reissner-Nordstr\"om black hole 
is more compact than a Schwarzschild one. 
  
\bigskip

\section{Thermodynamics in the TEGR and the standard first law of black hole mechanics}

$ $ 
 
In the standard formulation of the thermodynamics of black holes the gravitational energy is not taken into account 
and the internal energy of a black hole is considered to be given only by the black hole mass, which is parametrized 
in terms of its area, charge and angular momentum. Nevertheless, from the point 
of view of the conservation of energy and the thermodynamics, it is quite natural that gravitational energy should be taken into account if a 
black hole is to be considered as a thermodynamic system. That is, the internal energy of a black hole should be considered 
not only as its rest-mass and other, non-gravitational forms of energy (as electrostatic energy), 
but one should also take into account the gravitational energy as part of the total internal energy 
ascribed to a black hole. Also, taking into account the concept of gravitational pressure, it is 
natural to consider its role in black hole thermodynamics. As a result, in this section, we will analyze the role played by 
gravitational energy and gravitational pressure on the thermodynamics of a Reissner-Nordstr\"om black hole. Our aim in this 
section is basically to compute the quantity $dE + pdV$ and compare it with the standard first law of black hole dynamics. 
 
\bigskip
    
Let us first compute $pdV$. Since $\phi^{(r)1}$ is a density, the differential $pdV$ is evaluated as 

\begin{equation}
pdV=\biggl[ \int_S(-\phi^{(r)1})\,d\theta d\varphi \biggr]dr_+
=p\,dr_+\,,
\label{dif-vol}
\end{equation}
in which $S$ is the surface of constant radius $r=r_+$, with $p$ given by Eq. (\ref{pres-rad-hor-RN}). 
The differential $dr_+$ is obtained from Eq. (\ref{horizons}) and it reads

\begin{equation}
dr_{+} = {r_{+}\over \sqrt{M^{2} - Q^{2}}} \left( dM - {Q\over r_{+}}dQ\right)\,.
\label{dif-hor}
\end{equation}
It must be noted that as one is assuming that $dr_+$, $dM$ and $dQ$ are infinitesimals, the 
present analysis is not valid when $\sqrt{M^2-Q^2}$ approaches zero, \textit{i.e.},  when $Q$ is 
very close to $M$.

\bigskip

From Eq.'s (\ref{pres-rad-hor-RN}) and (\ref{dif-hor}) one is then left with

\begin{equation}
pdV= -\left( dM - {Q\over r_{+}}dQ\right)\,.
\label{dif-vol2}
\end{equation}

\bigskip

The differential $dE$ is easily obtained from Eq. (\ref{energia-hor-RN}) as

\begin{equation}
dE=dr_+ \,.
\label{dif-ener}
\end{equation}
Of course, this result also derives from the variation 

\begin{equation}
dE={{\partial E}\over {\partial M}}dM + {{\partial E}\over {\partial Q}}dQ\,, 
\end{equation}
as it should be.  

\bigskip

From (\ref{dif-vol2}) and (\ref{dif-ener}) we have

\begin{equation}
dE + pdV =\, \frac{M}{r_{+}}\,dr_{+} \,.
\label{1alei-tegr} 
\end{equation}
By replacing now Eq. (\ref{dif-hor}) into the equation above we are left with

\begin{equation}
dE + pdV ={M\over \sqrt{M^{2} - Q^{2}}}\left( dM - {Q\over r_{+}}dQ\right)\,.
\label{1alei-tegr2} 
\end{equation}

\bigskip

We will now come back to the expression given by Eq.(\ref{1alei-tegr}). Its right-hand side can be written in terms 
of the surface gravity of the black hole, {\it  i.e.}, in terms of

\begin{equation}
\kappa = \frac{\sqrt{M^{2}-Q^{2}}}{r_{+}^{2}}\,. 
\label{surf-grav}
\end{equation}

\bigskip

As the horizon area is $A=4\pi r_{+}^{2}$ it follows that Eq. (\ref{1alei-tegr}) can be rewritten as 

\begin{equation}
dE + pdV = \frac{1}{8\pi}\frac{M}{r_{+}^{2}}\,dA\,, 
\end{equation}
which, by virtue of Eq. (\ref{surf-grav}), can be written as

\begin{equation}
dE + pdV = \left(\frac{M}{\sqrt{M^{2}-Q^{2}}}\right)\frac{\kappa}{8\pi}\,dA\,. 
\label{tegr2}
\end{equation}
In particular, one sees that for a Schwarzschild black hole ($Q=0$), the latter result 
reduces to

\begin{equation}
dE + pdV = \frac{\kappa}{8\pi}\,dA\,. 
\label{tegr2-schwarz}
\end{equation}
Hence, for a Schwarzschild black hole the expression $dE + pdV$, obtained entirely in the context of the teleparallelism,  
coincides with the standard expression for the first law of black hole dynamics. 

\bigskip

Let us now compare the result given by Eq. (\ref{tegr2}) with the standard one, in what concerns a 
Reissner-Nordstr\"om black hole. We firstly recall that for the latter the first law of black hole 
dynamcis is given by

\begin{equation}
\frac{\kappa}{8\pi}\,dA = dM - \Phi _{H}dQ\,, 
\label{primeira-lei-mec-bn}
\end{equation}
where $\Phi_{H} = Q/r_{+}$ is the Coulombian potential at the black hole event horizon (the 
zero of the electric potential is taken at infinity). Considering now Eq. (\ref{tegr2}), and since $M\geq |Q|\,$, it 
follows that the factor which multiplies the term $(\kappa/8\pi)\,dA$ is greater than one. This implies 
that for a Reissner-Nordstr\"om black hole the following inequality holds 

\begin{equation}
dE + pdV > \frac{\kappa}{8\pi}\,dA\,. 
\label{tegr-desigualdade}
\end{equation}


\bigskip

If one now defines
\begin{equation}
TdS = dE + pdV 
\label{1a-lei-tds}
\end{equation}
as the first law of black hole dynamics, established entirely in the framework of the TEGR, 
it follows that Eq. (\ref{tegr2}) can be rewritten as 

\begin{equation}
TdS= \frac{M}{\sqrt{M^{2}-Q^{2}}}\,\left(\frac{\kappa}{8\pi}dA\right)\,. 
\label{TdS-tegr}
\end{equation} 
Although the area $A$ appears on the right-hand side of (\ref{TdS-tegr}), we stress it does not play any role in arriving 
at an expression for $dE + pdV (=TdS)$, but rather the latter is given by Eq. (\ref{1alei-tegr2}), which has been 
obtained without any need to resort to $A$. It is only after one arrives at an expression for $dE + pdV (=TdS)$ that it has been expressed, 
for convenience, in terms of $A$. We also remark that up to now we have not assumed that $S$ and $A$ are related by the 
standard, Bekenstein-Hawking relation. 

\bigskip

Assuming now that $T$ in Eq. (\ref{TdS-tegr}), which is the temperature of the black hole, is 
the Hawking temperature $\kappa /2\pi$, 
it follows from Eq. (\ref{TdS-tegr}) that 

\begin{equation}
dS= \frac{M}{\sqrt{M^{2}-Q^{2}}}\,\,\frac{dA}{4}\,. 
\label{T-tegr}
\end{equation}   
In this way, one is led to the result that, in the TEGR (due to both the gravitational energy and 
gravitational pressure so defined), the variation of the entropy of a Reissner-Nordstr\"om black hole is greater than the variation of the 
standard, Bekenstein-Hawking entropy, $S_{BH}=A/4$ (in natural units). For a Schwarzschild black hole ($Q=0$), 
the entropy (\ref{T-tegr}) so derived in the TEGR coincides with the standard one, even though the gravitational pressure is not null 
in this case. On the other hand, we recall that the Bekenstein-Hawking postulate, according 
to which the entropy of a black hole is given by the entropy-area relation $S_{BH}=A/4$, follows 
from the \textit{classical} laws of black hole dynamics together with the (quantum) Hawking temperature $\kappa /2\pi$. 
Hence, the result given by Eq. (\ref{T-tegr}) can be viewed as a possible modification 
of the entropy of a Reissner-Nordstr\"om black hole, as a result of considering both gravitational 
energy and gravitational pressure in formulating the classical laws of black hole dynamics. In this 
direction, as pointed out by York \cite{York-PRD1983}, we recall that the constant of proportionality
in the relation $S_{BH}=A/4$ was originally obtained from a mechanical-thermodynamical analogy based 
on the relation 

\begin{equation}
dM=\frac{\kappa}{8\pi}dA\,, 
\label{dM-dA} 
\end{equation}
which is derived from $M=(\kappa /8\pi)A$, which in its turn is valid for neutral 
nonrotating black holes.  It is only upon the identification by Hawking that (the black hole 
temperature is) $T=\kappa /2\pi$, it would follow from the hypothesis that (\ref{dM-dA}) 
can be written in thermodynamic form, with $dS_{BH}=(1/4)dA$, if and only if one assumes 
that the thermodynamic law for uncharged nonrotating black holes is given by
\begin{equation}
dM=TdS_{BH}\,,
\label{dM-dS} 
\end{equation}
from what one sees that there is no term corresponding to ``$pdV$" in the standard formulation of 
black hole thermodynamics. The point is that given $T=\kappa /2\pi$, the expressions for $M$ and $dM$ 
do not by themselves imply uniquely a value for the entropy. As York remarked \cite{York-PRD1983}, 
Eq. (\ref{dM-dS}) gives the simplest possibility that leads to an entropy-area relation.

\bigskip

\newpage

\section{Concluding Remarks}

$ $ 

The plausibility of a $pdV$ ``work" term in the first law of black hole thermodynamics is perhaps 
best summarized in the following remark by York \cite{York-PRD1983}: ``\textit{it is quite plausible that if ``heat" TdS 
is slowly added to a black hole in equilibrium, thereby causing it to expand, that it should do ``work" 
in lifting itself in its own gravitational `potential well'\,"}. Besides, as the electromagnetic field exerts 
pressure, one might expect that the gravitational field would behave in the same way. In fact, this has been 
shown to be to the case for gravitational waves \cite{Maluf-Ulhoa2008}. 

\bigskip

We have obtained the thermodynamic relation $TdS = dE + pdV$ (which is the first law of black hole thermodynamics) 
entirely within the framework of the TEGR, without identifying $dS$ with the variation $dA$ of the 
area of the event horizon of the black hole (see Eq. (\ref{1alei-tegr2})). However, in order to compare 
$TdS = dE + pdV$, as given by the TEGR result (\ref{1alei-tegr2}), with the (standard) $TdS_{BH}$, as given 
by the standard first law of black hole dyna-\linebreak mics (\ref{1a-lei-mec-bn}), we have written the right-hand 
side of Eq. (\ref{1alei-tegr}), which is a more compacted, preliminary form of Eq. (\ref{1alei-tegr2}), 
in terms of $(\kappa/8\pi)dA$. The result is given by Eq. (\ref{tegr2}), what implies that, in the framework 
of the TEGR, $TdS \geq (\kappa/8\pi)dA = TdS_{BH}$, where the inequality becomes an equality only for the 
particular case of a Schwarzschild black hole, whereas the inequality holds for a Reissner-Nordstr\"om black hole. 
This result imply that (i) for a Schwarzschild black hole the expression $dE + pdV$, obtained entirely in the 
context of the teleparallelism, coincides with the standard expression for the first law of black hole dynamics, 
while (ii) for a Reissner-Nordstr\"om black hole it leads to the fact that the efficiency of the Penrose process is 
less than in standard black hole thermodynamics. We note that the same conclusion has been achieved in 
the case of a Kerr black hole \cite{maluf-pressao-Kerr}.
      
\bigskip

The fact that the entropy given by Eq. (\ref{T-tegr}) (which has been derived entirely in the 
framework of the TEGR) is different from the standard, Bekenstein-Hawking entropy, $S_{BH}=A/4$, 
is not a surprise, of course. It should be noted that in the standard first law of black hole 
thermodynamics the role of internal energy is ascribed to the black hole mass, $M$, while in the TEGR 
it is played by the total energy enclosed by the event horizon of the black hole, $E$, which includes 
$M$ and other possible forms of energy. Also, the role of gravitational pressure, which is crucial in establishing in the 
TEGR the first law as $TdS = dE + pdV$, is a concept that is absent in the standard formulation of classical black hole 
dynamics. An exception is the consideration of a gravitational \textit{surface} pressure ascribed to the horizon of a 
black hole, as defined by Brown and York in the context of their quasi-local analysis \cite{Brown-York}. In this 
way, they arrive at the first law of black hole thermodynamics for a spherically symetric black hole, but with the black hole 
temperature blueshifted from infinity to a fixed distance $R$. Nevertheless, when the surface (of radius $R$) is taken as 
the horizon ($R=r_{+}$) the surface pressure diverges, as well as the temperature (see Eq.'s (6.19) and (6.20) of 
Ref. \cite{Brown-York}). In the context of the TEGR the gravitational pressure which enters into the first law of black hole 
dynamics is not a surface pressure, but a radial pressure over the event horizon, where it has a finite value 
(see Eq. (\ref{pres-rad-hor-RN})). Such a pressure is negative, what means it is directed towards the center 
of the black hole. Physically, one can view this as similar as in first law of ordinary thermodynamics, 
since corresponding to the fact that the Penrose process leads to the extraction of energy from the black hole, 
the black hole pressure is over the hoziron, (radially) directed to its center. In the case of a Kerr black 
hole the radial pressure over the event horizon is also negative, as shown in Ref. \cite{maluf-pressao-Kerr}, taking 
into account also the Penrose process. 

\bigskip

Since by no continuous process (such as the Penrose process) can the irreducible mass of a black hole 
be decreased (\textit{i.e.}, the inequality $dM^{2}_{irr}\geq 0$ holds), and as $A=16\pi M^{2}_{irr}$, it follows that 
$dA\geq 0$, that is, by no continuous process can the horizon area of a black hole be decreased \cite{Chandra}. 
Therefore, in view of Eqs. (\ref{tegr2-schwarz}) and (\ref{tegr-desigualdade}), one is led to $TdS=dE + pdV \geq 0$ 
\footnote{We recall that the equality in $TdS=dE + pdV \geq 0$ holds for a Schwarzschild black hole and that for the latter it is not possible 
to extract energy (mass) by the Penrose process, since $M_{irr}=M$, where $M$ is the mass of the black hole.}. 
This can be taken as the second law of black thermodynamics in the framework of the TEGR, though here it has been based on continuous 
processes involving only a single black hole. 

\bigskip

Finally, we note that the entropy-area relation given by Eq. (\ref{T-tegr}) can be considered ``holographic", 
analogously to the standard relation.  We tried to verify if a relation similar to Eq. (\ref{T-tegr}) holds for a 
Kerr black hole, by making use of the expression for $TdS$ obtained in Ref. \cite{maluf-pressao-Kerr}. However, 
in that case, the expression for the quantity $TdS=dE + pdV$ is not simple, but rather has a complicated form such 
that we could not make a conclusive statement. We hope to report about it in the future.

\bigskip

\textbf{Acknowledgements} 

\bigskip 

The authors are grateful to Professor J. W. Maluf for critical reading and providing comments on a earlier 
version of this manuscript, what has led to improvements of the presentation.


\begin{thebibliography}{10}
\bibitem{energ-KN-future}
This issue is under investigation by the present authors.
\end{thebibliography}
\end{document}